\documentclass[
reprint,
superscriptaddress,
 amsmath,amssymb,
 aps,prx
]{revtex4-2}

\usepackage{graphicx}
\usepackage{dcolumn}
\usepackage{bm}
\usepackage{color}
\usepackage{txfonts}

\newcommand{\delQ}{\nabla Q}

\newcommand{\QdelQ}{(\bm{Q}\nabla)\bm{Q}}

\begin{document}

\preprint{APS/123-QED}

\title{Experimental identification of force, velocity, and nematic order relationships in active nematic cell monolayers}

\author{Masahito Uwamichi}
\altaffiliation[Present Address: ]{Department of Basic Science, Graduate School of Arts and Sciences, The University of Tokyo, Tokyo, 113-0033, Japan}
\affiliation{Department of Physics, Graduate School of Science, The University of Tokyo, Tokyo, 113-0033, Japan}
\author{He Li}
\affiliation{Institute of Natural Sciences and School of Physics and Astronomy, Shanghai Jiao Tong University, Minhang District, Shanghai 200240, China}
\author{Zihui Zhao}
\affiliation{Institute of Natural Sciences and School of Physics and Astronomy, Shanghai Jiao Tong University, Minhang District, Shanghai 200240, China}
\author{Yisong Yao}
\affiliation{Institute of Natural Sciences and School of Physics and Astronomy, Shanghai Jiao Tong University, Minhang District, Shanghai 200240, China}
\author{Hideo Higuchi}
\affiliation{Department of Physics, Graduate School of Science, The University of Tokyo, Tokyo, 113-0033, Japan}
\affiliation{Universal Biology Institute, The University of Tokyo, Bunkyo-ku, Tokyo 113-0033, Japan}
\author{Kyogo Kawaguchi}
\affiliation{Nonequilibrium Physics of Living Matter RIKEN Hakubi Research Team, RIKEN Center for Biosystems Dynamics Research, 2-2-3 Minatojima-minamimachi, Chuo-ku, Kobe 650-0047, Japan}
\affiliation{RIKEN Cluster for Pioneering Research, 2-2-3 Minatojima-minamimachi, Chuo-ku, Kobe 650-0047, Japan}
\affiliation{
Institute for Physics of Intelligence, The University of Tokyo, 7-3-1 Hongo, Tokyo 113-0033, Japan}
\affiliation{Universal Biology Institute, The University of Tokyo, Bunkyo-ku, Tokyo 113-0033, Japan}
\author{Masaki Sano}
\affiliation{Institute of Natural Sciences and School of Physics and Astronomy, Shanghai Jiao Tong University, Minhang District, Shanghai 200240, China}
\affiliation{Universal Biology Institute, The University of Tokyo, Bunkyo-ku, Tokyo 113-0033, Japan}
\date{\today}

\begin{abstract}
Cell alignment often forms nematic order, which can lead to anomalous collective cell flow due to the so-called active force. Although it is appreciated that cell migration is driven by traction force, a quantitative evaluation of the relationships between the traction force, the nematic patterning, and the cell flow velocity is still elusive. Here we have found that cellular traction force aligns almost perfectly and is proportional in amplitude to the gradient of the nematic order tensor, not only near the topological defects but also globally. Furthermore, the flow in the monolayer was best described by adding nonlinear forces and a diffusion term derived from symmetry considerations. These nonlinear active forces enhance density instability but suppress bending instability, explaining why cell accumulation and dispersion can occur in neural progenitor cell culture while their ordering pattern is stable.
\end{abstract}

\maketitle

\textit{Introduction.---}
Multicellular tissue dynamics is fundamentally out of equilibrium since the components, the cells, are motile and can exert force on each other~\cite{marchetti2013,menon2010,ramaswamy2017,vicsek2012}. Recent experiments have opened up the possibility of utilizing multicellular dynamics as a platform to study nonequilibrium many-body physics. Cultured mammalian cells such as fibroblasts, myoblasts, neural progenitor cells, and even epithelial cells have been shown to exhibit features of active nematics, where cells forming liquid crystal-like order~\cite{kemkemer2000,duclos2014} are accompanied by rapid activity at the single-cell level~\cite{saw2017,kawaguchi2017}. These experiments have not only verified the ideas of active matter studies, but have also led to findings of unique features in real tissue dynamics\cite{doostmohammadi2015,duclos2018, yamauchi2020, comelles2021, copenhagen2021,maroudas-sacks2020a}.

One of the key concepts in the theory of active nematics is the role of the active force $\bm{F}^a$, which is a collective effect that emerges in the continuum description:
\begin{equation}\label{eq:delQ}
    \bm{F}^a = -\zeta \nabla\cdot \bm{Q},
\end{equation}
where $\zeta$ is the active stress parameter. Here, the nematic tensor order parameter $\bm{Q}$ and its derivative are defined by
\begin{equation}
Q (\bm{r}) = [Q_{ij}(\bm{r})] = \frac{1}{2}
\begin{bmatrix}
\langle \cos 2\theta \rangle_{\bm{r}} & \langle \sin 2\theta \rangle_{\bm{r}} \\
\langle \sin 2\theta \rangle_{\bm{r}} & -\langle \cos 2\theta \rangle_{\bm{r}}  
\end{bmatrix}
.
\end{equation}
and $\delQ:=\nabla \cdot{}\bm{Q}= \partial_j Q_{ij}$, with the position denoted as $\bm{r} = (x,y)$. Here $\theta(\bm{r})$ is the angle of the elongated axis of the cell relative to the $x$-axis, and $\langle \cdot \rangle_{\bm{r}}$ indicates the average taken at a mesoscopic scale that includes
multiple cells around ${\bm{r}}$. When the active stress parameter $\zeta$ is positive (negative), the system is called extensile (contractile). 
This force $\bm{F}^a$, which is absent in an equilibrium system, leads to characteristic features of active matter such as giant number fluctuation, long-range order, band chaos, and the anomalous dynamics around topological defects~\cite{ramaswamy2003,narayan2007,decamp2015,nishiguchi2017,kawaguchi2017,Guillamat,Hoffmann2022,Shankar2022}.
This active force [Eq.~(\ref{eq:delQ})] linearly proportional to the derivative of the nematic tensor can be derived by considering the continuum of a force dipole~\cite{aditisimha2002}, or more heuristically from symmetry arguments in continuum mechanics~\cite{ramaswamy2003}.

Although the traction force can be measured for collectively migrating cells, there has been no direct test of Eq.~(\ref{eq:delQ}) except for the case of a relatively low nematically ordered state in the epithelial sheet~\cite{nier2018} and model parameter fittings or machine learning without force measurement in bacterial colonies~\cite{li2019,copenhagen2021,Colen2021}.
Theoretical frameworks to interpret the relation between collective cell motion and intercellular stress have been reported~\cite{Blanch-Mercader2017,Notbohm2016,Zhang2023}.

Here, we have found that the traction forces are directed in the opposite direction to the active force predicted in Eq.~(\ref{eq:delQ}).
The linear relation between the traction force and $\delQ$ is consistent with known active defect dynamics \cite{kawaguchi2017, copenhagen2021} but with the opposite sign for extensile active nematics.
The instantaneous velocity of individual NPCs was along the principal direction of $\bm{Q}$, while the flow (time- and spatial-averaged velocity) were aligned with and partially proportional to $\delQ$. 
Furthermore, nonlinear active forces were found to have higher correlations with the cell flow. 
Combining these linear and nonlinear forces, we identify the time evolution equation for the cell flow that matches the experimental data.

\begin{figure*}[th]
\begin{center}
\includegraphics[width=17.8cm]{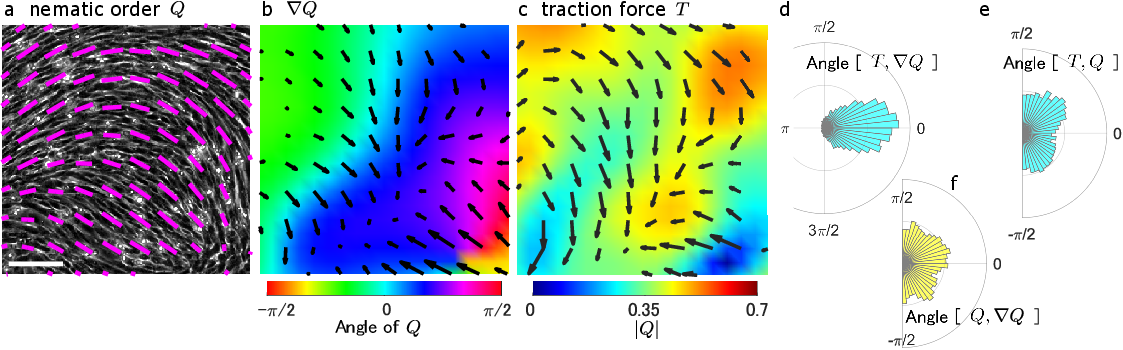}
\caption{\label{fig:QDT} 
Cell alignment pattern and traction force microscopy. (a) Phase contrast image and the nematic order in the monolayer of cultured neural progenitor cells. The scale bar indicates 100 ${\rm\mu{}m}$.
(b) Spatial derivative $\delQ:= \nabla \cdot \bm{Q}$ and (c) the traction force $\bm{T}$ overlaid on the magnitude of the local $\bm{Q}$. Magenta bars in (b) have lengths proportional to the eigenvalue and angle of the eigenvector. The polar histogram of the relative angles between (d) traction force and $\delQ$,  (e) $\pi$-symmetric angle of traction force, and (f) $\delQ$ relative to $\bm{Q}$.}
\end{center}
\end{figure*}

\begin{figure*}[tbph!]
 \begin{center}
\includegraphics{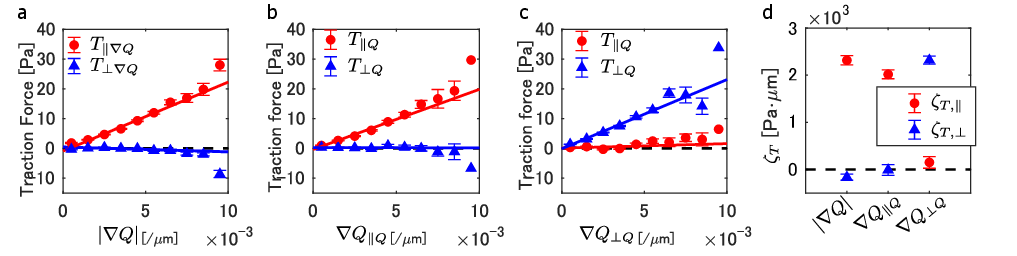}
\caption{\label{fig:zeta} Quantitative comparison between the alignment field and traction force. 
(a) The mean magnitude of traction force versus the binned magnitude of $\delQ$. Here, $T_{\parallel \delQ}$ and $T_{\perp \delQ}$, are the components of the traction force that are parallel and perpendicular to the direction of $\delQ$, respectively. 
(b-c) Traction force elements parallel $T_{\parallel Q}$ or perpendicular $T_{\perp Q}$ to $\bm{Q}$. Force was plotted against the elements of $\delQ$ that are parallel $\delQ_{\parallel Q}$ (b) or perpendicular $\delQ_{\perp Q}$ to $\bm{Q}$ (c), where we selected the angle of $\bm{Q}$ so that $\delQ_{\parallel Q} \geq 0$ (b) and $\delQ_{\perp Q} \geq 0$ (c) are the directions of $\bm{Q}$ upon projection. Symbols indicate the experimental results, solid lines show the result of fitting the traction force values as the first-order polynomial of $\delQ$ values (a-c). (d) The proportional coefficients obtained by fitting in (a-c), where $\zeta_{T,\parallel}$ is the coefficients of $T_{\parallel Q}$, $\zeta_{T,\perp}$ is the coefficients of $T_{\perp Q}$. Error bars indicate standard error (a-c) and 95\% confidence interval (d).}
 \end{center}
\end{figure*}

\textit{Comparing traction force and nematic pattern.---}
We performed traction force measurements for neural progenitor cells cultured on the soft silicone gel (see Supplementary Information).
The alignment field of the cells was quantified as the nematic tensor order parameter $\bm{Q}$, which was calculated from the intensity gradient of the phase contrast images using the structure tensor method~\cite{puspoki2016}.
The larger eigenvalue and the corresponding eigenvector of $\bm{Q}$ represent the strength and the orientation of the cellular alignment, respectively. The orientation of $\bm{Q}$ changed in a length scale much longer than the cell size or the window size for the calculation of $\bm{Q}$ [Fig.~\ref{fig:QDT}(a)], which is quantified by $\delQ$.

\begin{figure}[htbp]
\includegraphics[width=8.5cm]{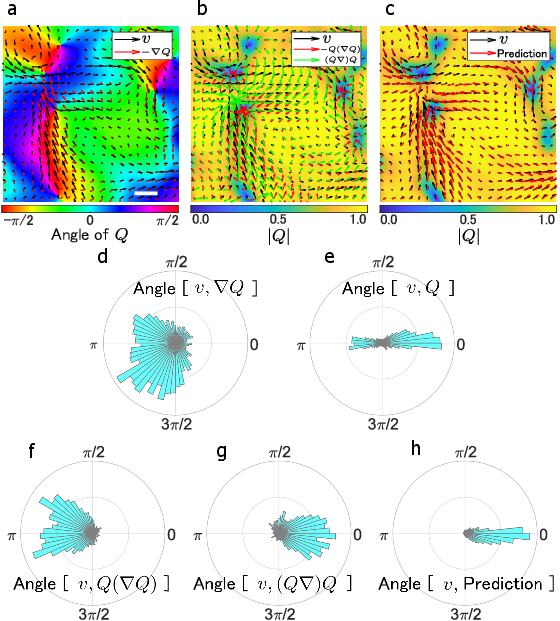}
\caption{\label{fig:QDV} Time-averaged velocity of NPCs. 
(a) $-\delQ$ and $\bm{v}$ vectors are overlaid on the colored orientation map of the nematic order parameter $\bm{Q}$. The scale bar indicates 100 ${\rm\mu{}m}$.
(b) Nonlinear forces $\bm{Q}\delQ$ and $\QdelQ$ are overlaid on the orientation map and compared with $\bm{v}$.  
(c) Predicted velocity and measured $\bm{v}$ were overlaid on the amplitude map of $|Q|$. 
Polar histograms of relative angles (d-i). 
(d) The angle of $\bm{v}$ relative to $\delQ$. 
(e) Angle between $\bm{v}$ and $\bm{Q}$.
(f) Angle of $\bm{Q}(\delQ)$ relative to $\bm{v}$. 
(g) Angle of $\QdelQ$ relative to $\bm{v}$. 
(h) Angle of predicted velocity relative to $\bm{v}$. }
\end{figure}

As a result of the TFM, we found that the traction force was well aligned with $\delQ$, but not with $\bm{Q}$ [Figs.~\ref{fig:QDT}(d,e)].
Orientations of $\bm{Q}$ are overlaid on the phase contrast image of cells in Fig.~\ref{fig:QDT}(a), and $\delQ$ vectors are displayed on the orientation field of $\bm{Q}$ in Fig.~\ref{fig:QDT}(b). Traction force vectors are displayed in Fig.~\ref{fig:QDT}(c).
The angle of traction force relative to $\delQ$ had a clear peak around zero degrees [Fig.~\ref{fig:QDT}(d)], while the traction force angle relative to $\bm{Q}$ was distributed almost isotropically [Fig.~\ref{fig:QDT}(e)]. The distribution of the angle of $\delQ$ relative to $\bm{Q}$ was also isotropic [Fig.~\ref{fig:QDT}(f)], suggesting that the direction of $\delQ$ is independent of the angle of alignment.

Next, we compared the amplitude of traction force with $\delQ$ [Fig.~\ref{fig:zeta}(a)]. 
To compare the two vector fields, we quantified the component of $\bm{T}$ in the direction parallel to and perpendicular to the local $\delQ$: $\bm{T} = (T_{\parallel \delQ}, T_{\perp \delQ})$.
We found that $T_{\parallel \delQ}$ is proportional to $|\delQ|$, while the average of $T_{\perp \delQ}$ is close to zero, indicating that $\bm{T} \simeq -\zeta_T \delQ$, where $\zeta_T<0$ is a scalar.

To further test whether the proportional relation between $\bm{T}$ and $\delQ$ is dependent on $\bm{Q}$, we quantified the components of $\bm{T}$ and $\delQ$ in the directions parallel and perpendicular with respect to the principal direction of  $\bm{Q}$:
\begin{eqnarray}
\bm{T} &=& (T_{\parallel Q}, T_{\perp Q}), \\
\delQ &=& (\delQ_{\parallel Q}, \delQ_{\perp Q}).
\end{eqnarray}
Note that the two vectors $\delQ_{\parallel Q}$ and $\delQ_{\perp Q}$ correspond to the two basic modes of 2D nematic order distortion, splaying and bending.
We found that each component in $\bm{T}$ and $\delQ$ is in linear relation [Figs.~\ref{fig:zeta}(b,c)]:
\begin{equation}
T_{\parallel Q} \propto \delQ_{\parallel Q},  \quad \quad T_{\perp Q} \propto \delQ_{\perp Q}.
\end{equation}
and the proportionality coefficients were similar to each other [Fig.~\ref{fig:zeta}(d)]. 
This indicates that the proportional relation holds between $\bm{T}$ and $\delQ$ irrespective of $\bm{Q}$, consistent with the independence between the angles of $\delQ$ and $\bm{Q}$ [Fig.~\ref{fig:QDT}(f)].
Thus, the macroscopic force exerted from the cells to the substrate is determined by the spatial gradient of the nematic order pattern, rather than the direction of cellular polarity or the principal angle of $\bm{Q}$.
The same result was obtained for a smooth-muscle cell line, SK-LMS-1, which also has a rod-like shape and exhibits a similar alignment pattern {\it in vivo} and {\it in vitro} \cite{Yang2010a, eskander2012} (Figs.~S5-S8 see Supplemental Information).

\begin{figure*}[th]
\includegraphics[width=17.8cm]{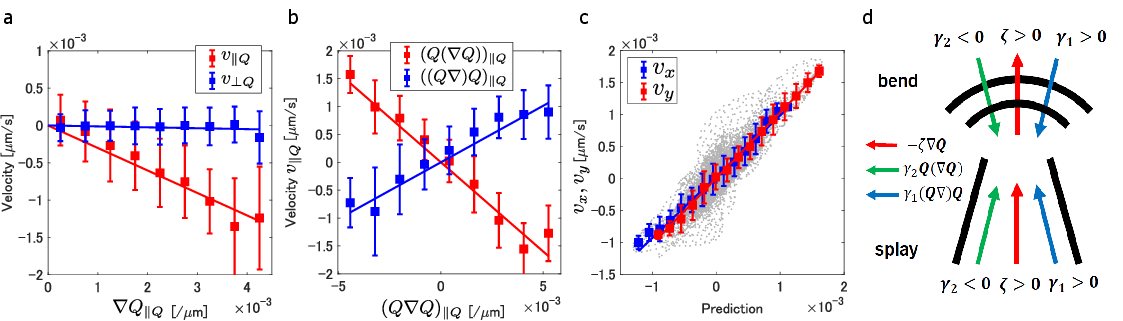}
\caption{\label{fig:gamma} The averaged magnitudes of $\bm{v}$ (vertical axes) are compared with the magnitudes of active forces and prediction (horizontal axes). Elements of $\bm{v}$ parallel and perpendicular to $Q$ were plotted against (a) $\delQ_{\parallel Q}$, (b) $(\bm{Q}(\delQ))_{\parallel Q}$ and $(\QdelQ)_{\parallel Q}$.
(c) $v_x$ and $v_y$ are compared with the prediction by Eq.(8).
Symbols denote experimental results, solid lines show the linear fitting. The scatter plot indicates raw data (gray). Error bars are standard deviations.
(d) Implication of active forces. Linear active force ($\zeta>0$) enhances bending perturbation, while the obtained parameters $\gamma{'}_1<0, \gamma{'}_2>0$ in Eq.~(8) stabilize the perturbation. For the splay perturbation, the same parameters do not alter the effect of the linear term. }
\end{figure*}

\textit{Predicting cell flow from ordering.---}
We next asked whether there is a relation between the collective force and the cell flow. Assuming that the nematic patterns vary slowly and that the cell flow is overdamped, we expect that the velocity of the net cell flow $\bm{v}$ should follow 
\begin{equation}\label{Eq:forcevelocity}
    \gamma \bm{v} \sim -\zeta \nabla \cdot \bm{Q},
\end{equation}
at the linear order on average. 
These assumptions can be validated as the speed of the change in nematically ordered pattern ($\sim 10^{-4} \mu \text{m/s}$) is small compared with the instantaneous velocity of individual cells ($\sim 10^{-2} \mu \text{m/s}$) and the cell flow ($\sim 10^{-3} \mu \text{m/s}$).

The cell flow measurement was performed for cultures on the glass substrate for the sake of long-term recording.
From this data, we tracked the cell nuclei to quantify the instantaneous velocity of each cell, and calculated the cell flow $\bm{v}$ by space- and time-averaging the velocity of cells within $32 \times 32 \ \mu\text{m}$ subgrids and 260 frames with 5 min intervals.

As shown in Fig.~\ref{fig:QDV}(a), the cell flow $\bm{v}$ was directed toward the opposite direction of $\delQ$, widely distributed in $[\pi/2, 3\pi/2]$ [Fig.~\ref{fig:QDV}(d)], whereas $\bm{v}$ was tightly bound along the principal axis of $\bm{Q}$ [Fig.~\ref{fig:QDV}(e)].
For a quantitative comparison, we decomposed $\bm{v}$ and $\delQ$ into the parallel and perpendicular components according to the principal axis of $\bm{Q}$ at each position. 
We found that the parallel components showed proportionality [Fig.~\ref{fig:gamma}(a)],
whereas the perpendicular component did not [Fig.~S10(b)].

Seeing that the simple linear relation between the cell flow $\bm{v}$ and $\delQ$ [Eq.~(\ref{Eq:forcevelocity})] does not hold, we next considered whether nonlinear terms involving $\bm{Q}$ can be utilized to predict the cell flow, as well as taking into account the diffusion term.
Nonlinear effects have been previously introduced as the anisotropic friction coefficient~\cite{kawaguchi2017}, and have also been discussed in \cite{Maitra2018}. Here we consider these factors from a symmetry argument.

The time evolution equations governing active nematics must satisfy the invariance by the transformation: $(x,y) \rightarrow (-x, -y)$, upon which $(\bm{v}, \bm{Q})$ change to $(-\bm{v}, \bm{Q})$. Taking the third order of ($\nabla, \bm{v}, \bm{Q}$)) leads to
\begin{eqnarray}
    \partial_t \rho\bm{v} &=& -\varGamma(\bm{Q})\rho\bm{v} -\zeta \delQ + \gamma_2 \bm{Q}(\delQ) + \gamma_1 \QdelQ  \nonumber \\
    &-& (\pi_0 + \lambda \bm{Q})\nabla \rho,
\end{eqnarray}
where $\rho$ is the cell density, $\varGamma(\bm{Q}) = \gamma_0 (\bm{I} - \epsilon \bm{Q})$ is the friction coefficient with $\epsilon$ being the anisotropy of the friction, and $I$ is the identity matrix. 
The nonlinear forces are defined as $\bm{Q}(\delQ):=Q_{ij}\partial_{k}Q_{jk}$ and $\QdelQ:=Q_{jk}\partial_{k}Q_{ij}$. 
Indeed, these vectors were pointing in a similar direction as the cell flow [Figs.~\ref{fig:QDV}(b,f,g)].
As for the parallel components according to $\bm{Q}$, these vectors also showed proportionality to the cell flow [Fig.~\ref{fig:gamma}(b)]. 
Note that nonlinear terms such as $v_j \partial_j v_i, v_i \partial_j v_j,...$ can be neglected in the overdamped approximation.
The diffusion term was also introduced earlier in the phenomenological model of dry active nematics~\cite{ramaswamy2003}.
The same equation can be derived with the Boltzmann-Ginzburg approach from a microscopic model of active nematics with repulsion~\cite{patelli2019}.

Assuming steady-state, we analyzed experimental data and identified the best-fitted parameters by using the Bayesian inference method for 
\begin{equation} \label{eq:velocity}
    (\bm{I} - \epsilon \bm{Q}) \bm{v} = -\zeta{'} \nabla{Q} + \gamma{'}_2 \bm{Q}(\delQ) + \gamma{'}_1 \QdelQ -(\pi_0{'}+\lambda{'}  \bm{Q})\nabla \rho,
\end{equation}
where we inferred the cell density $\rho$ from the fluorescence signal (see Supplementary Information, Fig.~S10(a)).
We used a Boussinesq-like approximation in fluid, i.e., we replaced $\rho$ with the mean density $\rho_0$ and ignored the density dependence except for $\nabla \rho$ term. The dynamics of cell density is governed by the continuity equation, $\partial_t \rho = -\partial_i (\rho v_i)$, together with Eq.~(\ref{eq:velocity}). For the coefficients defined as $\zeta{'} = \zeta/(\rho_0 \gamma_0), \gamma{'}_1 = \gamma_1/(\rho_0 \gamma_0), \gamma{'}_2 = \gamma_2/(\rho_0 \gamma_0), \pi_0{'} = \pi_0/(\rho_0 \gamma_0), \lambda{'}=\lambda/(\rho_0 \gamma_0)$,
we obtained the best-fitted results as $\epsilon=0.62(5), \zeta{'}=0.074(13), \gamma{'}_1=0.089(15), \gamma{'}_2=-0.029(24), \pi_0{'}=0.22(6), \lambda{'}=-0.068(34)$, where the errors are the standard deviation from the inference at three distinct areas in experiment.

The predicted cell flow by Eq.~(\ref{eq:velocity}) is shown and compared with experimental data in Figs.~\ref{fig:QDV}(c,h), and \ref{fig:gamma}(c).
The agreement is excellent except in the very vicinity of the core of the defect.
It is interesting to note that each term in Eq.~(\ref{eq:velocity})  only roughly aligns with the cell flow direction [Figs.~\ref{fig:QDV}(d,f,g)], whereas their linear combination can predict the cell flow quite accurately [Fig.~\ref{fig:gamma}(c)].

The implication of the linear and nonlinear active forces is illustrated in Fig.~\ref{fig:gamma}(d).
The linear active force with $\zeta>0$ enhances bending perturbation; the bending deformation should be amplified leading the system to undergo instability toward nematic chaos, which has not been observed in NPC experiments. 
On the other hand, the parameters for the nonlinear terms, observed to be $\gamma{'}_1>0$ and $\gamma{'}_2<0$ in Eq.~(\ref{eq:velocity}) as a result of the fitting,  stabilize the perturbation. 
The term corresponding to $\bm{Q}(\delQ)$ ($\gamma{'}_2<0$) has been introduced earlier to account for the stability of 2D active nematics~\cite{Maitra2018}. 
For the splay perturbation, the same parameters do not counter the effect of the linear term.

{\it Conclusion.---}
Here we have measured the traction force exerted by the neural progenitor cells to confirm its almost perfect alignment with the predicted linear active force, and further explained the cell flow by introducing nonlinear active forces. 
In our experiment, we could only observe the linear active forces in the traction force measurement. 
Since the estimated magnitude of the nonlinear term is comparable to the linear term, this result suggests that the nonlinear forces are not exerted as a traction force, and are rather emergent properties at the multicellular dynamics.

The relationship between traction force and the direction of cell flow has been discussed in terms of how contractile forces produce extensile ($\zeta>0$) behavior.
The velocity reversals or polar fluctuations of individual cell motion are thought to be sufficient for the extensile property~\cite{shi2013,Vafa2021,Killeen2022,patelli2019}.
Our result is consistent with a model that the single-cell level forces that are contractile (see Supplementary Material, Fig.~S9) lead to the migration of the cells, and that propelling motion adds up to an extensile time evolution with additional nonlinear terms at the collective level.
Further quantitative experiments of active nematics with force measurements will be important to seek the validity and mechanism of the nonlinear dynamics.

We thank Hugues Chaté and Hepeng Zhang for useful discussions.
We acknowledge financial support from National Natural Science
Foundation of China Grant Numbers 12174254 and No. 12250710131 (to M.S), JSPS KAKENHI Grant Numbers JP18K13515, JP18H04760, JP19H05795,	JP19H05275, JP21H01007, and JP23H00095 (to K.K.).

\bibliography{main}

\end{document}


\preprint{APS/123-QED}

\title{Supplementary Materials for ``Experimental identification of force, velocity, nematic order relationships in active nematic cell monolayers"}

\author{Masahito Uwamichi}
\altaffiliation[Present Address: ]{Department of Basic Science, Graduate School of Arts and Sciences, The University of Tokyo, Tokyo, 113-0033, Japan}
\affiliation{Department of Physics, Graduate School of Science, The University of Tokyo, Tokyo, 113-0033, Japan}
\author{He Li}
\affiliation{Institute of Natural Sciences and School of Physics and Astronomy, Shanghai Jiao Tong University, Minhang District, Shanghai 200240, China}
\author{Zihui Zhao}
\affiliation{Institute of Natural Sciences and School of Physics and Astronomy, Shanghai Jiao Tong University, Minhang District, Shanghai 200240, China}
\author{Yisong Yao}
\affiliation{Institute of Natural Sciences and School of Physics and Astronomy, Shanghai Jiao Tong University, Minhang District, Shanghai 200240, China}
\author{Hideo Higuchi}
\affiliation{Department of Physics, Graduate School of Science, The University of Tokyo, Tokyo, 113-0033, Japan}
\affiliation{Universal Biology Institute, The University of Tokyo, Bunkyo-ku, Tokyo 113-0033, Japan}
\author{Kyogo Kawaguchi}
\affiliation{Nonequilibrium Physics of Living Matter RIKEN Hakubi Research Team, RIKEN Center for Biosystems Dynamics Research, 2-2-3 Minatojima-minamimachi, Chuo-ku, Kobe 650-0047, Japan}
\affiliation{RIKEN Cluster for Pioneering Research, 2-2-3 Minatojima-minamimachi, Chuo-ku, Kobe 650-0047, Japan}
\affiliation{
Institute for Physics of Intelligence, The University of Tokyo, 7-3-1 Hongo, Tokyo 113-0033, Japan}
\affiliation{Universal Biology Institute, The University of Tokyo, Bunkyo-ku, Tokyo 113-0033, Japan}
\author{Masaki Sano}
\affiliation{Institute of Natural Sciences and School of Physics and Astronomy, Shanghai Jiao Tong University, Minhang District, Shanghai 200240, China}
\affiliation{Universal Biology Institute, The University of Tokyo, Bunkyo-ku, Tokyo 113-0033, Japan}

\date{\today}
\maketitle

\section{\label{sec:method}Materials and Methods}

\subsection*{Materials}
\begin{itemize}[itemsep=0pt, topsep=0pt]
    \item Neural progenitor cells (NPCs) were a kind gift from Ryoichiro Kageyama. 
    \item SK-LMS-1 cells were obtained from ATCC (HTB-88). 
    \item DMEM/F-12 (Invitrogen 11330)
    \item EGF (Wako 53003-018)
    \item bFGF (Wako 060-04543)
    \item N-2 Plus media supplement (N2plus; R\&D AR003)
    \item Laminin (Wako 120-05751)
    \item DMEM (Nacalai 08458)
    \item FBS (Gibco 26140079)
    \item Penicillin/streptomycin (P/S; Invitrogen 15140122)
    \item PBS (Wako 14249-95)
    \item Accutase (ICT 12679-54)
    \item 35mm plastic dish (Falcon 35-3001)
    \item glass base dish (Iwaki 3960-035)
    \item Cy52-276 (Dow Corning)
    \item 3-Aminopropyltriethoxysilane (APTS; Sigma 440140),
    \item 1-Ethyl-3-(3-dimethylaminopropyl)-carbodiimide (EDC; Dojindo W001)
    \item HEPES (Gibco 15630080)
    \item FluoSpheres ($\phi$0.2$~\mu$m, carboxylated, 660/690 nm, Invitrogen F8807)
    \item Stainless steel microspheres ($\phi$300-330 $\mu$m, Cospheric SSMMS-7.8)
\end{itemize}

\subsection*{Gel substrate for traction force microscopy (TFM)}
The silicone gel substrates for TFM were manufactured as follows. First, the glass base dishes were activated by applying air plasma. The activated dishes were incubated with silane-coupling solution (1\% APTS in milliQ water) for 5 min, and subsequently with the diluted FluoSpheres suspension (0.3-0.9\% beads, 0.1\% EDC and 20 mM HEPES in milliQ water) for 10 min. The mixture of Cy52-276 A and B components was spin-coated and cured on the glass base dishes. Finally, the gel surface was activated and covered with FluoSpheres, by the protocol used for the glass base. The gel elasticity was roughly controlled by changing the A/B ratio (0.90 to 1.40 in weight) of Cy52-276 mixture. We used A/B = 1.18 (Young's modulus $E=5$ kPa) for collective NPCs, 0.90 ($E=1\times10^2$ kPa) for collective SK-LMS-1 cells, or 1.40 ($E=1$ kPa) for both isolated NPCs and isolated SK-LMS-1 cells.

\subsection*{Cell culture}
NPCs were cultured in DMEM/F-12 supplemented with EGF, bFGF, N2plus, and P/S. The substrates used for NPCs were coated with laminin, by means of adding laminin stock solution (1.2\% v/v) into the growth medium. SK-LMS-1 cells were cultured in DMEM mixed with FBS and P/S. Plastic dishes were used for +1/2 defect tracking experiments, glass base dishes for cell flow measurements, or the gel substrates for TFM. For the experiments on collective cells, imaging was started after the substrate got fully covered with cells. For TFM with isolated cells, smaller number of the cells were seeded, and we took images within 1 or 2 days.

\subsection*{Imaging}
For TFM, we constructed an imaging system based on IX70 microscope body. Fluorescence of the cell nuclei or the beads was respectively excited by 532 nm or 641 nm laser, while Kohler illuminating was used for phase contrast images. All the images were taken by sCMOS camera (zyla 5.5; 2048$\times$2048 pixels) which was synchronized with the shutters for the light sources and the piezo actuator attached to the objective lens (Olympus, 20x, NA 0.45). 

We took timelapse pictures with 1 frame/min for at least 1 hour in each TFM imaging. The total number of field of views was 25 for 3 samples of collective NPCs, or 10 for 1 sample of collective SK-LMS-1 cells. At each time frame, 4 ${\rm\mu}$m around both glass and substrate surface was imaged by moving the objective for 1 ${\rm\mu}$m/step to select best focal plane at image analysis [Fig.~\ref{fig:NPCcolPIV}(c-f) green].

After imaging the gel substrate with cells, the cells were removed by applying Accutase for 10 min. Accutase and the remaining cells were washed out with water, then the dish was filled with fresh growth medium. The reference images of the gel without cells were taken one hour after the washing [Fig.~\ref{fig:NPCcolPIV}(c-f) red].

Finally,  for the gel stiffness measurement, stainless beads were gently put on the gel substrate and fluorescent images and phase contrast images were taken. After taking phase contrast images at the best focal plane of stainless beads, fluorescent images around the glass and gel surface were taken by moving the objective for 1 ${\rm\mu}$m/step so that all the fluorescent beads were included. At least five stainless beads were imaged for every sample.

\subsection*{Image analysis}
The alignment field of the cells was extracted from phase contrast images, utilizing the structure tensor method as previously shown \cite{puspoki2016,kawaguchi2017}. In brief, the structure tensor $\bm{J}$, which is the second moment of the image intensity gradient was calculated;
\begin{equation}
    \bm{J} = \left[ \begin{array}{cc}
                      J_{xx}  &  J_{xy} \\
                      J_{xy}  &  J_{yy}
                    \end{array} \right]
      \equiv \left[ \begin{array}{cc}
                \langle I_x^2\rangle_w & \langle I_x I_y\rangle_w \\
                \langle I_x I_y\rangle_w & \langle I_y^2 \rangle_w
                \end{array}   \right] ,
\end{equation}
where $I_x$ and $I_{y}$ are the $x$ and $y$ components of the intensity gradient, and $\langle \cdot \rangle_w$ indicates 2D-Gaussian filtering with standard deviation $w=52~{\rm\mu}$m for NPCs or $w=67~{\rm\mu}$m for SK-LMS-1 cells. The eigenvalues $\lambda_{max} > \lambda_{min}$ and the corresponding eigenvectors $\bm{e}_{max}$, $\bm{e}_{min}$ of $\bm{J}$ were calculated, and the angle of $\bm{e}_{min}$ was extracted as the local alignment angle $\theta$. The strength $q$ of the alignment was estimated from the coherence of the structure tensor anisotropy.
\begin{eqnarray}
    q &=& \frac{\lambda_{max} - \lambda_{min}}{\lambda_{max} + \lambda_{min}}\nonumber\\
	  &=& \frac{\sqrt{(J_{yy} - J_{xx})^2 + 4J_{xy}}}{J_{xx} + J_{yy}} .
\end{eqnarray}
Using $\theta$ and $q$, the $Q$-tensor was constructed as
\begin{equation}
    \bm{Q}(\bm{r},t) = q(\bm{r},t) \left[\begin{array}{cc}
       \cos2\theta(\bm{r},t)  & \sin2\theta(\bm{r},t) \\
        \sin2\theta(\bm{r},t) & -\cos2\theta(\bm{r},t)
    \end{array}  \right].
\end{equation}

\begin{figure*}[p]
\includegraphics{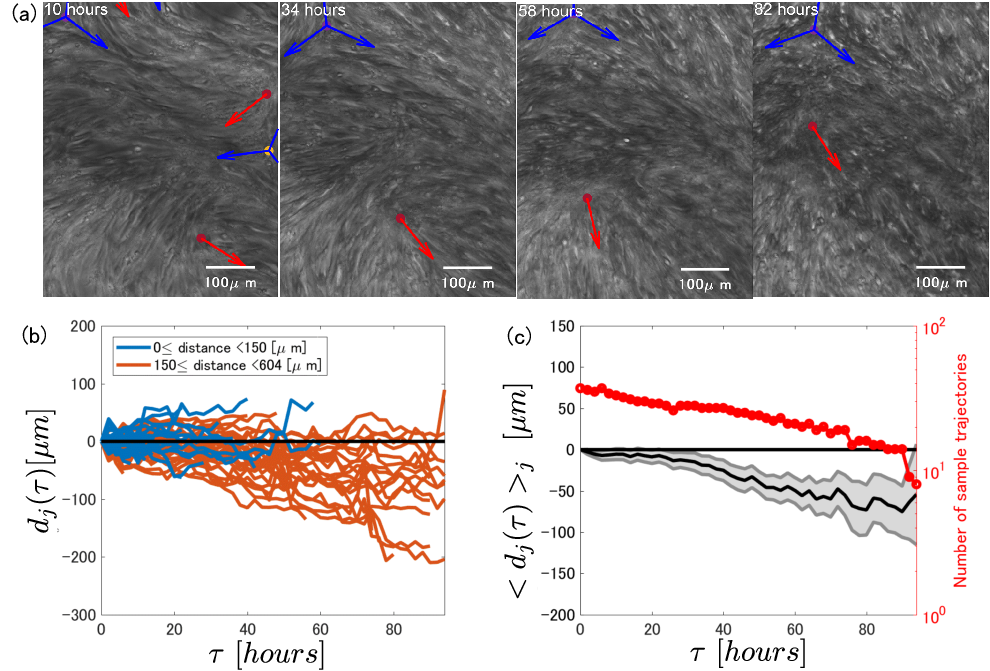}
\caption{\label{fig:NPCdefect} (a) +1/2 topological defect found in the NPCs culture. The red quiver indicates the comet's tail direction. (b) Displacement of +1/2 defects toward the comet tail direction. Blue curves indicate the defects that have the nearest neighbor defect within 150 $\mu$m, and orange curves show the isolated defects. (c) The ensemble average of the tail-directed displacement for the isolated +1/2 defects (b). The black curve indicates the ensemble average and the gray area shows a 95\% confidence interval of the average. Red points indicate the number of trajectories included in the average.}
\end{figure*}

\begin{figure*}[p]
\includegraphics{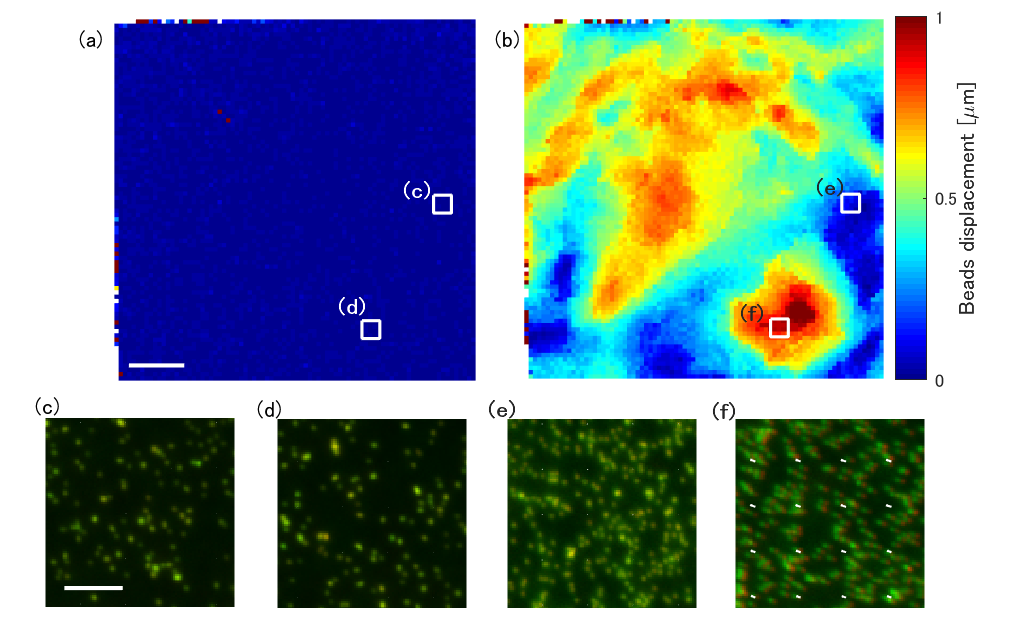}
\caption{\label{fig:NPCcolPIV} Result of PIV for collective NPCs traction force microscopy. (a-b) Amplitude of the calculated displacement of the beads bound on the glass (a) or on the gel surface (b). (c-f) Fluorescent images of the region are indicated by white squares in (a) and (b). Images taken before (green) and after (red) the removal of the cells were merged. Calculated displacement is indicated by white quivers, though it is hard to see in (c-e). Scale bars indicate 100 $\mu$m (a-b) and 10 $\mu$m (b), respectively.}
\end{figure*}

For the tracking of topological defects, the defects were detected by calculating the winding number $\xi$ defined by
\begin{equation}\label{eq:winding}
    \xi(\bm{r}, t) = \frac{1}{2\pi} \int^{2\pi}_0 \frac{ d\theta(\bm{r}+\delta \bm{e}_{\phi}, t)}{d\phi} d\phi ,
\end{equation}
following the procedure used in \cite{Decamp2015}.
Here, $\phi$ is the argument of the polar coordinates centered on the position $\bm{r}$. As $\xi$ is discrete, islands with nonzero $\xi$ were easily segmented, and the geometric center $\bm{r}^{(d)}_j$ of the $j$-th island was assigned as the position of the $j$-th defect. 
At the same time, the ‘comet tail’ direction $\phi_j$ of the $j$-th defect is calculated by fitting [Fig.~\ref{fig:NPCdefect}(a),~\ref{fig:SKLMS1defect}(a,b)]:
\begin{equation}
    \theta(\bm{r}^{(d)}_{j}+\delta \bm{e}_{\phi}, t) - \phi_j = \xi_j (\phi - \phi_j) .
\end{equation}

The detected +1/2 defects were tracked by nearest-neighbor tracking. The velocity $\bm{v}^{(d)}_{j}$ of $j$-th defect was calculated as $\bm{v}^{(d)}_{j} = d\bm{r}^{(d)}_{j}/dt$. To compare the dynamics of +1/2 defects with their morphology, we calculated the displacement $d_{j}(\tau)$ of the defects toward their own tail direction;
\begin{equation}
    d_{j}(\tau) = \int^{t^j + \tau}_{t^j} dt \bm{v}^{(d)}_{j}(t) \cdot \bm{e}^{{\rm tail}}_j(t) ,
\end{equation}
where $\bm{e}^{{\rm tail}}_j$ is the unit vector toward the tail direction $\phi_j$, $t^j$ is the initial frame, and $\tau$ is the size of the time window [Fig.~\ref{fig:NPCdefect}(b,c),~\ref{fig:SKLMS1defect}(c)].

\subsection*{Particle image velocimetry (PIV)}
To quantify the gel deformation, the displacement of the gel-top beads was calculated by PIV. 
In advance, horizontal drift was eliminated by the global registration between the images of beads bound on the glass, using {\it imregtform} function in MATLAB. 
By the registration, an affine transformation, which encodes translation and rotation in the horizontal plane, was calculated for each of the glass-bound beads images, by minimizing the mean square difference from the corresponding reference image. 
The drift correction was done by applying the extracted affine transformation to each gel-bound bead image [Fig.~\ref{fig:NPCcolPIV}(a)].

After the drift correction, the bead displacement field was calculated by mPIV, an open-source PIV calculation code for MATLAB [Fig.~\ref{fig:NPCcolPIV}(b)]. In PIV, the local translation $\bm{u}$ was calculated by maximizing the cross-correlation between the small window of the gel-bound beads image and that of the corresponding reference image;
\begin{eqnarray}
    \bm{u}&&(\bm{r},t) = \nonumber\\ &&\argmax_{\bm{u}^{\prime}} \left[ \sum_{|\bm{k}|\neq0} \tilde{I}(\bm{k}; W\{\bm{r}\},t) \overline{\tilde{I}_{ref}(\bm{k}; W\{\bm{r}+\bm{u}^{\prime}\})} \right] ,
\end{eqnarray}
where $W\{\bm{R}\}$ is the square ROI with $24 \times 24$ pixels $^2$ ($7.8 \times 7.8~{\rm \mu m}^2 $), centered at the position $\bm{R}$. $\tilde{I}(\bm{k};W\{\bm{R}\},t)$, $\tilde{I}_{ref}(\bm{k};W\{\bm{R}\})$ are the Fourier transform with wave number $\bm{k}$ of the intensity of the $W\{\bm{R}\}$ for the beads image on the frame $t$ and on the reference frame, respectively.

\subsection*{Gel stiffness measurement by indentation}
\begin{figure*}[p]
\includegraphics{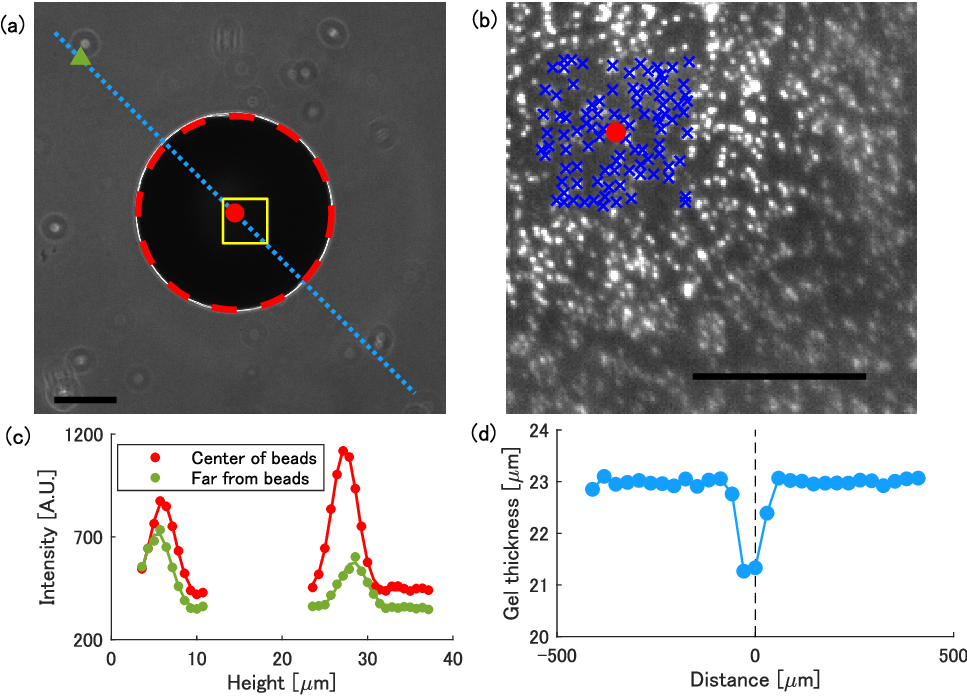}
\caption{\label{fig:indentation}  Gel stiffness measurement for $A/B=1.18$. (a) Phase contrast image of stainless steel beads overlaid with the detected center (red circle) and radius (red broken line). (b) Magnified image of fluorescent beads on the gel surface, in the region indicated by yellow square in (a). Blue crosses indicate the local maxima around the center of the bead, detected by ImageJ. (c) Fluorescence intensity averaged for the detected local maxima, plotted as a function of the height. Red circles are the values obtained from the local maxima around the center of the beads, and green is that around the point indicated by the green triangle in (a). Solid lines indicate the results of Gaussian fitting. (d) Gel thickness is calculated along the blue dotted line in (a). The broken line is the guide for the center of the beads. Scale bars (black solid line) indicate (a) 100 $\mu$m or (b) 30 $\mu$m, respectively. }
\end{figure*}

Elasticity E of the gel substrate was measured from the indentation depth measured when stainless steel beads were put on the gel \cite{Frey2007}. First, according to Hertz theory for a gel with infinite thickness, the elasticity $E_{\rm Hertz}$ of the substrate can be calculated as
\begin{eqnarray}
	E_{\rm Hertz} &=& \frac{3(1-\nu^2)F}{4R^{1/2}\delta^{3/2}} ,\\
	F &=& \frac{4\pi}{3} R^3 (\rho_{\rm steel} - \rho_{\rm medium}) g ,
\end{eqnarray}
where $\nu$ is the Poisson’s ratio of the gel, $F$ is the gravitational force caused by the difference between the density of the medium $\rho_{\rm medium}$ and of the steel beads $\rho_{\rm steel}$, $R$ is the radius of the steel beads, $\delta$ is the indentation depth, $g$ is the gravitational acceleration. Considering the finite thickness $h$ of the gel, the correction $E/E_{\rm Hertz}$ has been approximated in \cite{Long2011} as
\begin{eqnarray}
	\frac{E}{E_{\rm Hertz}} &=& \frac{1 + 2.3w}{1 + 1.15w^{1/3} + \alpha(\frac{R}{h})w + \beta(\frac{R}{h})w^2} ,\\
	w &=& \left(\frac{R\delta}{h^2}\right)^{3/2} ,\\
    \alpha \left(\frac{R}{h} \right) &=& 10.05 - 0.63 \sqrt{\frac{h}{R}} \left(3.1 + \left(\frac{h}{R}\right)^2 \right) ,\\
    \beta \left(\frac{R}{h} \right) &=& 4.8 - 4.23 \left (\frac{h}{R} \right )^2 .
\end{eqnarray}
Here, stainless steel beads with density $\rho_{\rm steel} = 7.8\times10^3~{\rm kg/m^3}$ were used. We set $\nu = 0.5$, $\rho_{\rm medium} = 1.0\times10^3~{\rm kg/m^3}$, $g = 9.8~{\rm m/s^2}$. $R = \sqrt{A/\pi}$ was calculated from the cross-section area $A$ of stainless beads was measured by applying {\it Particle Analysis} on the phase contrast image thresholded with {\it Make Binary} in ImageJ [Fig.~\ref{fig:indentation} (a)]. 

To estimate the gel thickness, we analyzed the images of fluorescent beads as follows. 
First, the horizontal position $\bm{r}_i$ of the $i$-th beads was determined as the intensity local maxima at the focal plane [Fig.~\ref{fig:indentation} (b)]. 
Using the positions, we calculated the intensity profile $I^{b} (z)$ averaged for the beads detected in a certain area by
\begin{equation}\label{eq:beadHeight}
 I^{b} (z) = \langle I(\bm{r}_i; z) \rangle _i,
\end{equation}
where $I(\bm{r}; z)$ is the fluorescence intensity [Fig.~\ref{fig:indentation} (c)]. 
By fitting $I^{b} (b)$ with a Gaussian function, we obtained the height of the beads. Finally, Gel thickness $h$ was measured as the height of the gel-bound beads relative to the glass-bound beads, at a distance of $\sim350~{\rm {\mu}m }$ from the center of the steel beads[Fig.~\ref{fig:indentation} (d)]. Indentation depth $\delta$ was calculated as $\delta = h - h^{\prime}$, where $h^{\prime}$ is the gel thickness at the center of the steel beads. For the region of averaging at Eq.~\ref{eq:beadHeight}, we used the window size of $\sim 26 \times 26 \ {\rm {\mu}m}^2$.

\subsection*{Traction force microscopy}
The traction force field was calculated from the obtained gel deformation, considering the finite gel thickness $h$ \cite{Trepat2009}. In brief, the constitutive equation of the linear elastic theory and the boundary conditions of the symmetric stress tensor $\sigma$, the deformation rate $\epsilon$, and the deformation $\bm{u}$ can be written as follows;
\begin{eqnarray}
	\sigma_{\alpha\beta} &=& \frac{E}{1+\nu} \left[\epsilon_{\alpha\beta} + \frac{\nu}{1-2\nu} \sum_{\gamma}\epsilon_{\gamma\gamma}\delta_{\alpha\beta}\right] ,\\
	\epsilon_{\alpha\beta} &=& \frac{1}{2} \left[\partial_{\alpha} u_{\beta} +\partial_{\beta} u_{\alpha}\right] ,\\
	\bm{u} (x,y; z=h) &=& \bm{u}^{PIV}(x,y) ,\\
	\bm{u} (x,y; z=0) &=& 0 ,\\
	\sigma_{zz} (x,y; z=h) &=& 0 ,
\end{eqnarray}
where, $\bm{u}^{PIV}$ indicates the result of PIV, and $\alpha$, $\beta$, and $\gamma$ indicate the horizontal coordinates $x$, $y$ or the height $z$ relative to the glass-gel interface. In the Fourier space $\bm{k}=(k_x, k_y)$, the solution on the Fourier transformed traction force $\tilde{T}_{\alpha} = \tilde{\sigma}_{z\alpha} |_{z=h}$ is;
\begin{eqnarray}
	\tilde{T}_{\alpha} (k_x,k_y) &=& \sum_{\beta} \tilde{G}_{\alpha\beta} (k_x,k_y) \tilde{u}_{\beta} (k_x,k_y) ,\\
	\tilde{\bm{G}} (k_x, k_y) &=& \frac{E}{2(1+\nu)k} \left[\frac{c}{s} \left(\begin{array}{cc}
	   k_y^2  & -k_x k_y \\
	   -k_x k_y  & k_x^2
	\end{array} \right) \right. \nonumber\\
	&&+ \frac{\Gamma}{1-\nu} \left. \left( \begin{array}{cc}
	   k_x^2  & k_x k_y \\
	   k_x k_y  & k_y^2
	\end{array} \right) \right] ,\\
	\Gamma &=& \frac{(3-4\nu)c^2 +(1-2\nu)^2 +(kh)^2}{(3-4\nu)sc + kh},
\end{eqnarray}
where $k = |\bm{k}|$, $c = \cosh(kh)$, $s = \sinh(kh)$.

Since $\tilde{\bm{G}}$ increases as $k$ increases, the traction force solution becomes overfitted.
To avoid this, the traction force was smoothened by the Tikhonov regularization \cite{Sabass2008,Han2015,Plotnikov2014}. 
Tikhonov regularization is based on Bayesian inference assuming Gaussian distribution both for the noise included in the PIV result and for the amplitude of traction force, which leads to
\begin{eqnarray}
	\tilde{\bm{T}} (\bm{k}; \lambda) = \argmin_{\tilde{\bm{T}}^\prime} \left[ | \tilde{\bm{u}}(\bm{k}) - \tilde{\bm{G}}(\bm{k})\tilde{\bm{T}^\prime} |^2 + \lambda^2 | \tilde{\bm{T}^\prime} |^2 ]\right],~ ~
\end{eqnarray}
where $\lambda$ is the ratio between the standard deviations of PIV noise and of traction force amplitude. $\lambda$ was determined by a heuristic method called Lcurve \cite{Hansen1993}, in which the point with the largest curvature was chosen from the curve drawn on the $(\log(\eta),\log(\rho))$ space, where the signal $\eta$ and the noise $\rho$ are
\begin{eqnarray}
    \eta(\lambda) &=& \sum_{\bm{k}} |\tilde{\bm{T}} (\bm{k}; \lambda)|^2 ,\\
    \rho(\lambda) &=& \sum_{\bm{k}} | \tilde{\bm{u}}(\bm{k}) - \tilde{\bm{G}}(\bm{k})\tilde{\bm{T}} (\bm{k}; \lambda)|^2.
\end{eqnarray}

\subsection*{Estimation of the parameter set in Eq.~(8) in the main article}

Since the measurement of $\partial_t \bm{V}$ suffers from large measurement noise, we applied the Bayesian inference method in order to estimate the parameters in  Eq.~(8) to predict $\bm{v}$.
In this formalism, the presence of a term including $\bm{v}$ in the right-hand side of Eq.~(8) may cause trouble due to a strong correlation with the L.H.S. of Eq.~(8). 
However, this term appears multiplied by $\bm{Q}$, which is in a random direction with $\bm{v}$, thus after averaging it will vanish.

The parameter vector $\bm{p}$ and observation vector $\bm{V}$ are defined by
\begin{eqnarray}
\bm{p} &=& (\epsilon, \zeta, \gamma_2, \gamma_1, \lambda, \pi_0)^T , \\
\bm{V} &=& (v_x^1, v_x^2,\cdots,v_x^N, v_y^1,v_y^2,\cdots,v_y^N)^T ,
\end{eqnarray}
where the superscript $i=1,2,...,N$ denotes the observation at each grid point $i$, $N$ is the total number of observation points (typically $N = 65 \times 65 = 4225$ in one cropped image).

Now, by letting 
\begin{eqnarray}
\bm{q}^i_{j} =  \begin{pmatrix}
 (Q^iv^i)_j \\
 (\nabla Q^i)_j \\
 (Q^i\nabla Q^i)_j \\
 (Q^i\nabla)Q^i_j \\
 Q^i\partial_j \rho^i \\
 \partial_j \rho^i 
\end{pmatrix}^T,
\end{eqnarray}
for $j \in (x,y)$, the input matrix $\bm{X}$ is defined by
\begin{eqnarray}
\bm{X} &=& 
\begin{pmatrix}
\bm{q}^1_{x}  \\
\vdots  \\
\bm{q}^N_{x}   \\
\bm{q}^1_{y}  \\
\vdots  \\
\bm{q}^N_{y}   \\
\end{pmatrix}.
\end{eqnarray}

The Bayesian inference reformulates the equation into the problem of maximizing the posterior probability distribution function $\Pi(\bm{p}|\bm{\bm{V}})$ of the parameter set for the given observation velocity data:
\begin{equation}
    \Pi(\bm{p}|\bm{\bm{V}}) = L(\bm{V}|\bm{p}) \times \pi(\bm{p}),  
\end{equation}
where $L(\bm{V}|\bm{p})$ is the likelihood functional and $\pi(\bm{p})$ is the prior distribution of the parameter set. 
We assume that these are both Gaussian:
\begin{eqnarray}
    L(\bm{V}|\bm{p}) &=& \left(\frac{1}{\sqrt{2\pi \sigma^2}}\right)^{2N} \exp\left[ -\frac{||\bm{V}-\bm{X}\bm{p}||^2}{2\sigma^2}\right] ,\\
    \pi(\bm{p}) &=& \left(\frac{1}{\sqrt{2\pi\sigma_0^2}}\right)^{5} \exp\left[-\frac{\bm{p}_0^t\bm{p}_0}{2\sigma_0^2}\right] \nonumber \\
    &\times& \left(\frac{1}{\sqrt{2\pi\sigma_{\pi_0}^2}}\right) \exp\left[-\frac{(\pi_0 - \bar{\pi}_0)^2}{2\sigma_{\pi_0}^2}\right] ,
\end{eqnarray}
where the parameter vector is separated into two parts, $\bm{p_0} = (\epsilon, \zeta, \gamma_2, \gamma_1, \lambda)$ and $\pi_0$.
Here, $\pi_0$ is the diffusion constant of the density. This is validated if we substitute Eq.~(7) into the equation of continuity, $\partial_t \rho = -\partial_j (\rho v_j)$. The diffusion constant needs to be positive for the stability of the continuity equation. Thus, we assume that $\pi_0$ obeys Gaussian distribution with a positive mean ($\bar{\pi}_0 > 0$) and a variance $\sigma_{\pi_0}^2$ in the prior distribution.

Maximizing the log-likelihood functional $\log (\Pi(\bm{p}|\bm{\bm{v}}))$ corresponds to minimizing $S$, which is defined by
\begin{equation}
    \text{min}_{\bm{p}} S = \text{min}_{\bm{p}} [(\bm{V}-\bm{X} \bm{p})^t (\bm{V}-\bm{X} \bm{p}) +\Lambda \bm{p}_0^t \bm{p}_0 + \Lambda_{\pi_0} (\pi_0 - \bar{\pi}_0)^2] ,
\end{equation}
where $\Lambda = \sigma^2/\sigma_0^2$ and $\Lambda_{\pi_0} = \sigma^2/\sigma_{\pi_0}$.
The minimization condition leads to
\begin{equation}
    (\bm{X}^t \bm{X} + \Lambda \delta_{jk} + (\Lambda_{\pi_0}-\Lambda) \delta_{j,6}) \bm{p}  = \bm{X}^t \bm{V}  + \Lambda_{\pi_0}\bar{\pi}_0 \delta_{j,6} .
\end{equation}
The optimal parameter set is then obtained as 
\begin{equation}
    \bm{p} = (\bm{X}^t \bm{X} + \Lambda \delta_{jk} + (\Lambda_{\pi_0}-\Lambda) \delta_{j,6} )^{-1} (\bm{X}^t \bm{V} + \Lambda_{\pi_0}\bar{\pi}_0 \delta_{j,6})  .
\end{equation}

The variances of prediction error $\sigma^2$ and the magnitude of parameters $\sigma_0^2$ are initially set as $\sigma^2 = 10^{-6}$, $\sigma_0^2 = 10^{-4}$. The initial $\sigma_{\pi_0}$ selected by the estimate of the diffusion constant: $\pi_0 = D = v_0^2/\alpha$ where $v_0$ is the instantaneous velocity of the cell and $\alpha$ is the velocity reversal rate. We also used $v_0 = 0.01~{\rm \mu m/s}$ and $\alpha = 1/3600~{\rm s^{-1}}$. Using the calculated $\bm{p}$, $\sigma^2$ and $\sigma_0^2$ were updated by
\begin{eqnarray}
    \sigma^2 &=& \frac{1}{2N} ||\bm{v}- \bm{X}\bm{p}||^2 ,\\
    \sigma_0^2 &=& \frac{1}{5}\bm{p_0}^t\bm{p}_0,
\end{eqnarray}
which were subsequently used to update $\bm{p}$. This cycle was iterated until the relative change of the prediction error per iteration becomes smaller than $10^{-3}$.

The fitting results depend on the number of parameters. When we employ only the linear active force term $-\zeta \nabla \bm{Q}$ and $\nabla \rho$ terms, the estimation error is large. So far, the best fit was obtained by the Eq.~(8). The second choice would be neglecting the anisotropic diffusion term, by setting $\lambda'=0$. This choice of parameters give a slightly more scattered result compared with Fig.~4 as shown in Fig.~\ref{fig:Lambda_zero}.

\subsection*{Tracking cell nuclei} 
To obtain the cellular velocity, cell nuclei were tracked by {\it TrackMate}, an ImageJ plugin, in the fluorescent images. In this experiment, we used a stable NPC line expressing nucleus markers (H2B-mCherry) \cite{kawaguchi2017}.

For $i$-th trajectory ${\bm r}_i (t)$, velocity ${\bm v}_i (t)$ was calculated in each time interval of $dt=5~\text{min}$:
${\bm v}_i (t) = \left({\bm r}_i(t+dt) - {\bm r}_i(t)\right)/dt$. 
The net flow field ${\bm v}({\bm r})$ was obtained by taking the temporal and ensemble average of the velocities around each grid point ${\bm r}$ with an interval of 26~$\mu$m;
\begin{equation}
	{\bm v}({\bm r}) = \langle {\bm v}_i(t) \rangle _{(i, t) \in nn({\bm r})}~,
\end{equation}
where $\langle\cdot\rangle_{(i, t) \in nn({\bm r})}$ means average over sets of $(i,t)$ such that the nearest grid point of ${\bm r}_i (t)$ is ${\bm r}$. 

\subsection*{Estimation of density and $\bm{Q}$ in Eq.~(8)}
In estimating the parameters of Eq.~(8), density $\rho(\bm{r})$ and $\bm{Q}(\bm{r})$ are required. In order to estimate $\rho(\bm{r})$, we used fluorescent images of nuclei tagged by mCherry were used. We averaged 200 frames with a time interval of 5 min over 16.7 hours. After averaging frames,
a Gaussian filter with $\sigma=32.5~\mu$m  was applied to obtain final $\rho(\bm{r})$ [Fig.~\ref{fig:rho}(a)]. The validity of using fluorescent images in estimating the density of the cells was examined by counting the number of nuclei in the same grid-size regions and we confirmed good proportionality between the two methods.
In the Bayesian inference method to Eq.~(8), estimation of $\bm{Q}(\bm{r})$ patterns were performed as follows. The aforementioned structure tensor method was used to obtain the local orientation of cells. 
The intensity gradient $I_x$ and $I_{y}$ were calculated at pixel level ($\Delta x =0.65~{\rm\mu}$m), then 2D-Gaussian filtering with standard deviation $w=16.25~{\rm\mu}$m was used to obtain $\langle I_x^2\rangle_w, \langle I_y^2 \rangle_w$ and $\langle I_x I_y\rangle_w$.
The local orientation of the pattern was calculated by
\begin{equation}
   \theta = - \frac{1}{2}\arctan \left(\frac{2\langle I_x I_y\rangle_w}{\langle I_y^2 \rangle_w-\langle I_x^2\rangle_w} \right) .
\end{equation}
The nematic tensor order parameter $\bm{Q}$ and scalar order parameter $S$ were calculated using the formula:
\begin{equation}
Q (\bm{r}) = \frac{1}{2}
\begin{bmatrix}
\langle \cos 2\theta \rangle_{\bm{r}} & \langle \sin 2\theta \rangle_{\bm{r}} \\
\langle \sin 2\theta \rangle_{\bm{r}} & -\langle \cos 2\theta \rangle_{\bm{r}}  
\end{bmatrix}
= 
\begin{bmatrix}
Q_{xx} & Q_{xy} \\
Q_{xy} & -Q_{xx}   
\end{bmatrix}
.
\end{equation}
and $S(\bm{r})$ is defined by $\sqrt{Q_{xx}^2(\bm{r}) + Q_{xy}^2(\bm{r})}$, where the average $\langle \cdot \rangle_{\bm{r}}$ is calculated with the grid size of $32.5~{\rm\mu}$m. 

\subsection*{Comparison between vectors and tensors}

To investigate the alignment and magnitude correlation among a set of vectors, we adopted a technique that involves decomposing a given vector into components parallel or perpendicular to another vector at each position $\bm{r}$. This other vector is termed the reference vector. When dealing with a vector $\bm{A}(\bm{r})$, the angle of the reference vector, denoted as $\theta_{A}(\bm{r})$, is simply the angle of $\bm{A}$ itself. If $\bm{A}(\bm{r})$ represents a tensor, $\theta_{A}(\bm{r})$ is defined as the angle of the eigenvector corresponding to the tensor's largest eigenvalue.

Once $\theta_{A}(\bm{r})$ is determined, we decompose the vector $\bm{B}(\bm{r})$ into parallel ($\bm{B}_{\parallel{}A}$) and perpendicular ($\bm{B}_{\perp{}A}$) components with respect to the reference vector:
\begin{eqnarray}
B_{\parallel{}A} &=& \bm{B}\cdot\bm{e}_{\theta_{A}}, \\
B_{\perp{}A} &=& \bm{B}\cdot\bm{e}_{\theta_{A}+\pi/2},
\end{eqnarray}
where $\bm{e}_{\phi}$ is a unit vector directed at angle $\phi$ in the laboratory frame.
The components of $\bm{B}$ are compared to the magnitude of $\bm{A}$ to examine if they are proportional, and the angle between vectors $\bm{B}$ and $\bm{A}$ is calculated as $\theta_{B} - \theta_{A}$.

\section{\label{sec:SKresult}Comparing traction force and nematic pattern in SK-LMS-1 cells}
We performed traction force microscopy for the SK-LMS-1 cells, a human cancer smooth-muscle cell line from leiomyosarcoma. As already implied in morphological observation of leiomyosarcoma and SK-LMS-1 cells \cite{Yang2010a, eskander2012}, cells have an elongated shape and are aligned with each other [Fig.~\ref{fig:SKLMS1_QT}(a)], with half-integer topological defects indicating that the system is nematic. 
We found that the topological defects moved against the comet tail direction [Fig.~\ref{fig:SKLMS1defect}], similar to that of NPCs [Fig.~\ref{fig:NPCdefect}], indicating that the dynamics is extensile. 

The Q-tensor and its divergence $\nabla Q$ were calculated with the same method as in NPCs [Figs.~\ref{fig:SKLMS1_QT}(a,b)]. $\bm{Q}$ showed a much longer length scale of change than cellular length and the scale of calculation. To note, since the cells are larger, we calculated $\bm{Q}$ and $\nabla Q$ in a slightly longer length scale than that used for NPCs.

SK-LMS-1 cells were cultured on the soft gel substrate in order to measure the traction force. Here, we used a stiffer gel than that used for NPCs since the deformation induced by SK-LMS-1 was larger. Traction force vectors are displayed in Fig.~\ref{fig:SKLMS1_QT}(c). We found that the measured traction force was aligned with $\nabla Q$ [Fig.~\ref{fig:SKLMS1_rose}(a)]. In this case, the traction force also showed alignment with $\bm{Q}$, but the angles of the traction force relative to $\bm{Q}$ distributed more broadly than those relative to $\nabla Q$ [Fig.~\ref{fig:SKLMS1_rose}(b)]. As $\nabla Q$ similarly aligned with $\bm{Q}$ [Fig.~\ref{fig:SKLMS1_rose}(c)], the cell sheet likely had more splay pattern than bending pattern.

Next, we compared the magnitude of traction force with $\nabla Q$. The traction force vector $\bm{T}$ was decomposed into components $T_{\parallel\nabla Q}$ and $T_{\perp\nabla Q}$, which are parallel or perpendicular to $\nabla Q$. Comparing each component to $|\nabla Q|$, only $T_{\parallel\nabla Q}$ showed positive proportionality, whereas $T_{\perp\nabla Q}$ did not [Fig.~\ref{fig:SKLMS1_zeta}(a)].

Furthermore, we checked whether the proportionality of traction force to $\nabla Q$ is dependent on $\bm{Q}$. Parallel and perpendicular components of both $\bm{T}$ and $\nabla Q$ were calculated according to the direction of $\bm{Q}$, and were compared. Here, both $T_{\parallel Q}$ and $T_{\perp Q}$ were respectively linear to the counterpart $\nabla Q_{\parallel Q}$ and $\nabla Q_{\perp Q}$ [Figs.~\ref{fig:SKLMS1_zeta}(b,c)]. The proportional coefficients were consistent over the comparisons [Fig.~\ref{fig:SKLMS1_zeta}(d)], thus the linear relationship between traction force and $\nabla Q$ did not depend on $\bm{Q}$. 

To summarize, these results suggest that traction force was related to $\nabla Q$ rather than to $\bm{Q}$. The alignment of traction force and $\bm{Q}$ is likely due to the splay modes in the sheet, where $\nabla Q$ and $\bm{Q}$ align. The relation between traction force and $\nabla Q$ is qualitatively similar to that observed in NPCs, nevertheless, the proportional coefficients are different in two orders of magnitude. This suggests the active force is conserved among nematically ordered cell sheets.

\section{Comparison between the force dipoles of isolated and collective cells}
\begin{figure*}[p]
\includegraphics{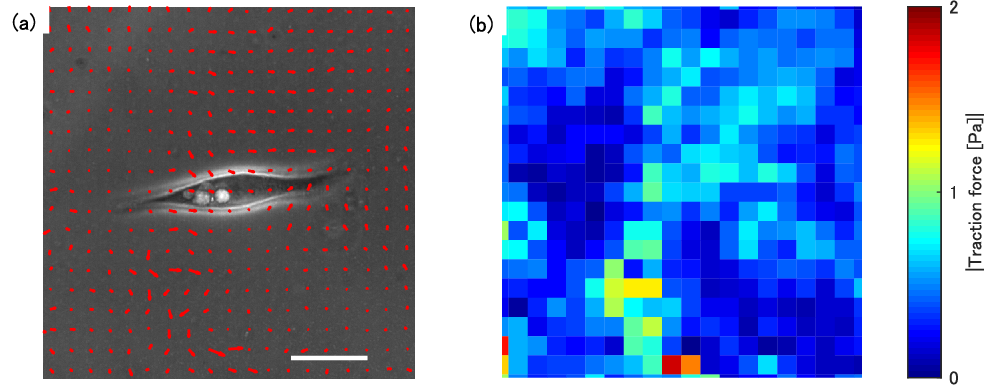}
\caption{\label{fig:singleNPCPIV}  Traction force measurement for the isolated NPCs. (a) The calculated traction force (red arrows) overlaid on the image of cell. (b) Magnitude of traction force. Scale bar indicates 20 $\mu$m.}
\end{figure*}

\begin{figure*}[hp]
\includegraphics{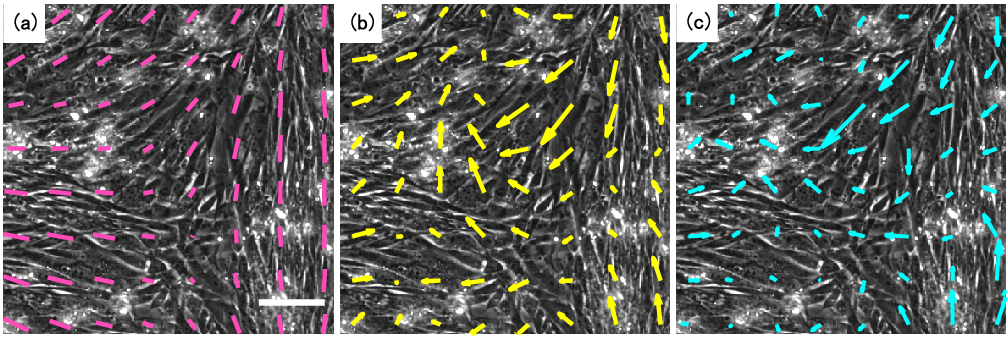}
\caption{\label{fig:SKLMS1_QT} Cell alignment pattern and traction force microscopy in SK-LMS-1 cell sheet. $\bm{Q}$ (a), $\nabla Q$ (b), and the traction force (c) measured for the collective SK-LMS-1 cell culture overlaid on the image of cells.  Scale bar indicates 100 $\mu$m.}
\end{figure*}

\begin{figure*}[p]
\includegraphics{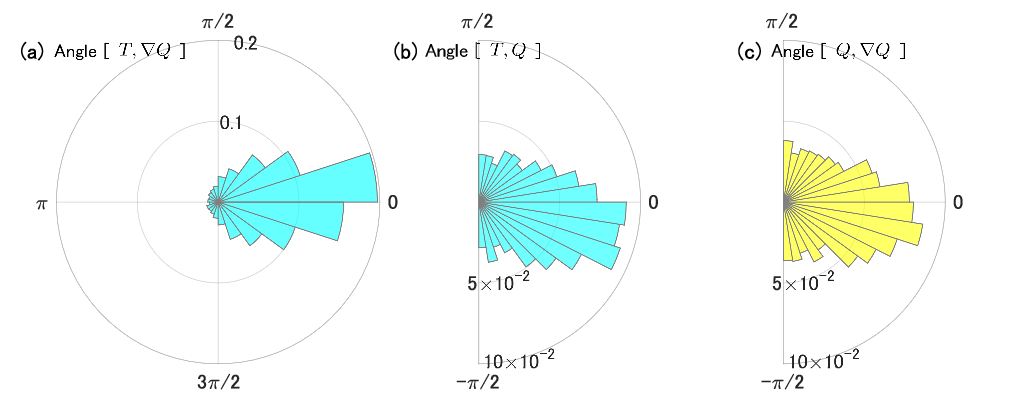}
\caption{\label{fig:SKLMS1_rose} Angular correlation between $\bm{T}$, $\nabla Q$ and $\bm{Q}$ of SK-LMS-1 cells on gel substrate. (a-b) Histogram of the angle of $\bm{T}$ relative to $\nabla Q$ (a) or $\bm{Q}$ (b). (c) Histogram of the angle of $\nabla Q$ relative to $\bm{Q}$.}
\end{figure*}

\begin{figure*}[p]
\includegraphics{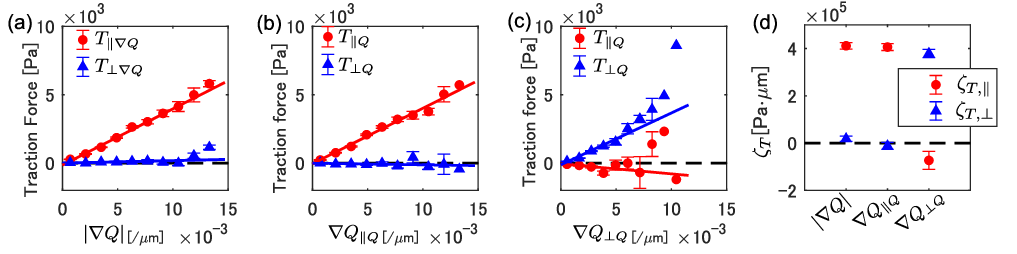}
\caption{\label{fig:SKLMS1_zeta}  The averaged magnitude of traction force binned with magnitude of $\nabla Q$ in the collective SK-LMS-1 case. (a) Traction force was divided into two elements parallel $T_\parallel$ or perpendicular $T_\perp$ to $\nabla Q$ and representatively plotted against $|\nabla Q|$. (b-c) Traction force elements parallel $T_{\parallel Q}$ or perpendicular $T_{\perp Q}$ to $\bm{Q}$. Force was plotted against $\nabla Q$ elements parallel $\nabla Q_{\parallel Q}$ (b) or perpendicular $\nabla Q_{\perp Q}$ to $\bm{Q}$ (c). Here we selected the angle of $\bm{Q}$ {\it s.t.} $\nabla Q_{\parallel Q} \geq 0$ (b) or $\nabla Q_{\perp Q} \geq 0$ (c) as the direction of $\bm{Q}$ when projection. Symbols indicate the experimental results, solid lines show the result of fitting the traction force values as the first order polynomial of $\nabla Q$ values (a-c). (d) The proportional coefficients obtained by fitting in (a-c), where $\zeta_{T,\parallel}$ is the coefficients of $T_{\parallel\nabla Q}$ or $T_{\parallel Q}$, $\zeta_{T,\perp}$ is that of $T_{\perp\nabla Q}$ or $T_{\perp Q}$ . Error bars indicate standard error (a-c) and 95\% confidence interval (d).}
\end{figure*}

\begin{figure*}[p]
\includegraphics{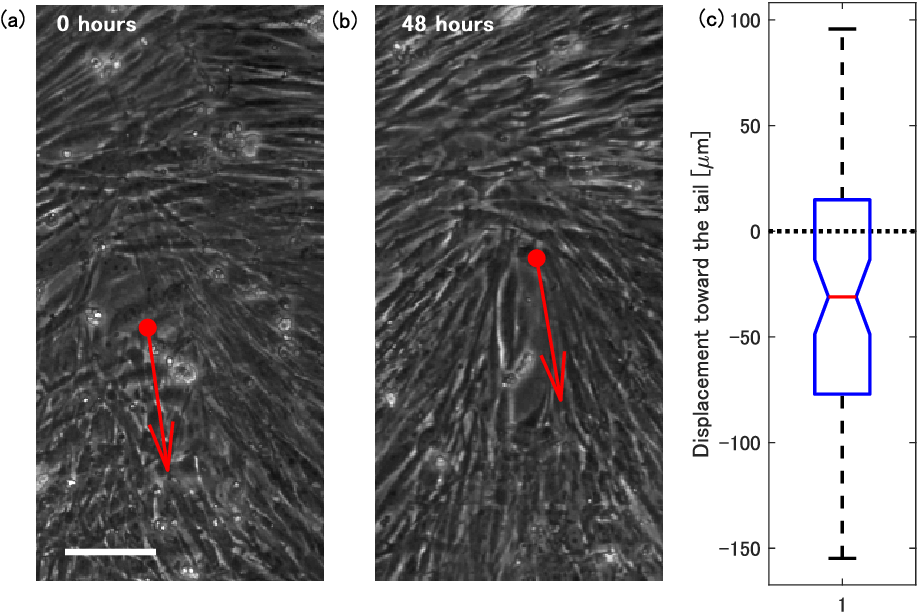}
\caption{\label{fig:SKLMS1defect}  Tracking result of the +1/2 defects of SK-LMS-1 cells. (a,b) Phase contrast images of the cell culture (a) at the start of observation or (b) 48 hours later. Red point and quiver show the location and the tail direction of the detected +1/2 defect, respectively. (c) Box-whisker plot of the displacement of +1/2 defects projected on the tail direction. Median is indicated by red line, 25\% and 75\% by blue lines at the ends of the box, 0\% and 100\% by black solid lines at the ends of whiskers. Notch indicates the 95\% confidence interval of the median. Dotted line is a guide for zero. Scale bar indicates 100 $\mu$m.}
\end{figure*}

By assuming that the active stress~\cite{hatwalne2004} is the sum of the force dipoles that the isolated NPCs have shown in the previous study \cite{Shi2009}, we can estimate $\zeta$.
The theory considers an elongated cell as a force dipole with strength $lf$, where $l$ is the length of the longer axis, $f$ is the force generated at both ends of the cell. As $\zeta$ is the force dipole density, $\zeta=lf\phi$ holds for the cases of 2D culture with cell density $\phi$. Here, $f$ is smaller than $1~{\rm nN}$ in \cite{Shi2009}, and we used the empirical values $l\leq100~{\rm\mu m}$, $\phi\sim 10^3~{\rm mm^{-2}}$ coming from typical confluent cell number $10^6$ for a 35 mm-diameter culture dish. From these parameters, we obtain $\zeta \sim1\times10^2 {\rm~Pa\cdot\mu m}$.
The fact that this $\zeta$ is much smaller than the value ($\zeta \sim2\times10^3 {\rm~Pa\cdot\mu m}$) measured in our traction force experiment for collective cells indicates that the force is not just the simple sum of individuals, and rather there is a collective effect. We note that we could not measure the traction force of isolated NPCs in our experiments, which is likely due to the force being too small [Fig.~\ref{fig:singleNPCPIV}].

\begin{figure*}[p]
\includegraphics{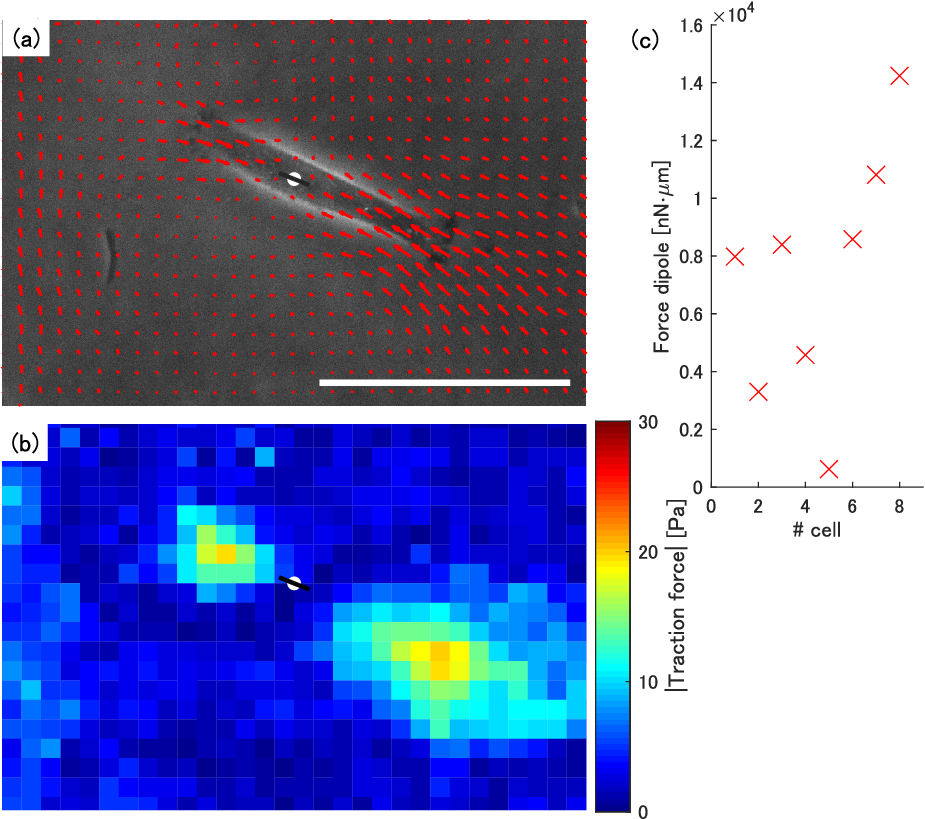}
\caption{\label{fig:singleSKTFM}  Traction force microscopy of the isolated SK-LMS-1 cells. (a) Traction force field (red quivers) superimposed on the phase contrast image of an isolated SK-LMS-1 cell. The center (white circle) and the direction of longer axis (black bar) of the cell are respectively indicated. Scale bar indicates 100 ${\rm\mu}$m. (b) Color plot of the magnitude of the traction force at the same area as (a). (c) Force dipole calculated for each cell (N=8).}
\end{figure*}

\begin{figure*}[p]
\includegraphics[width=8.5cm]{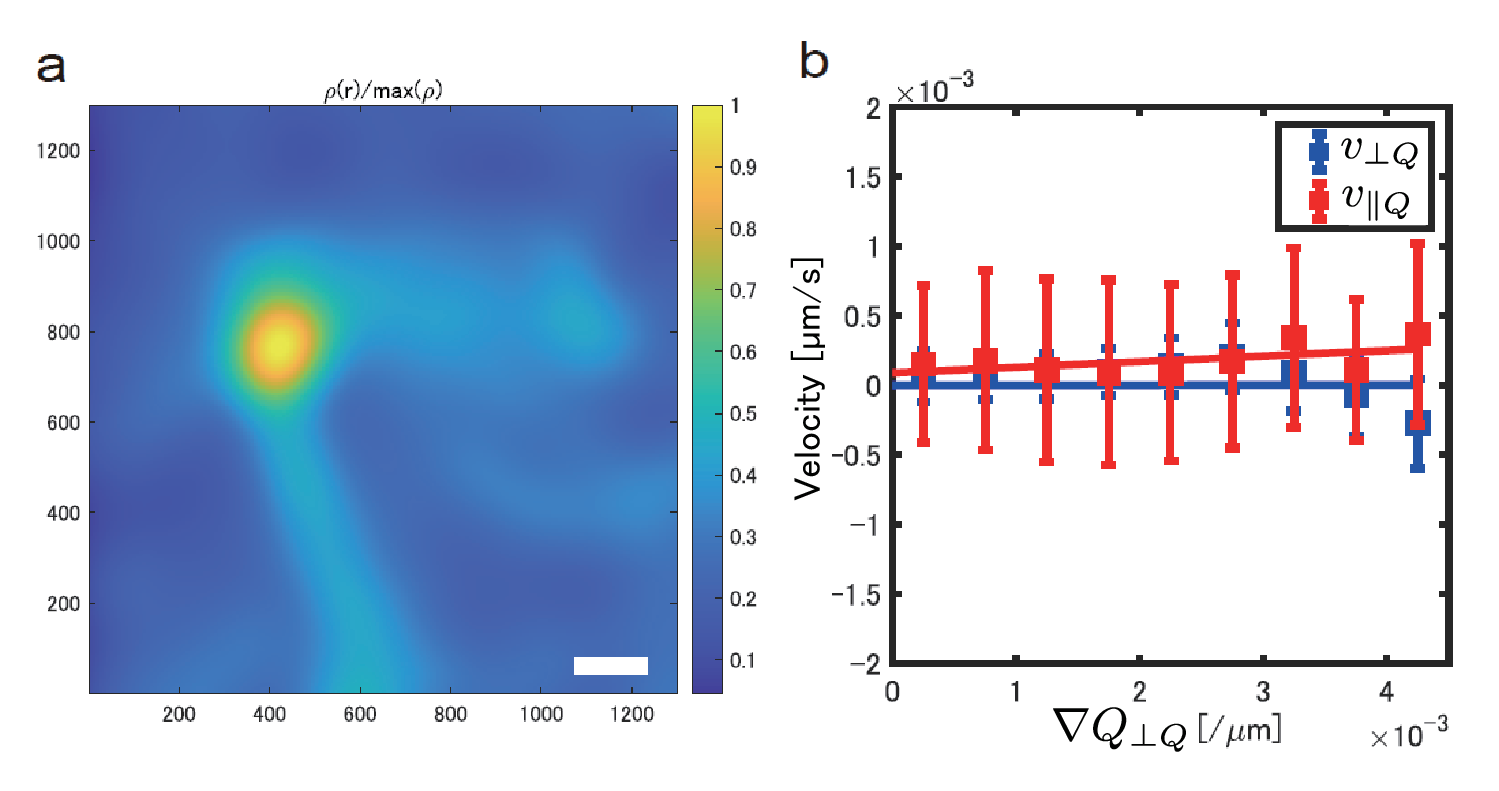}
\caption{\label{fig:rho}  a. Density profile estimated by averaging fluorescent images for 16.6 hours (time interval: 5 min). Scale bar indicates 100 $\mu$m. b. Supplementary data for Fig.4. Elements of $\bm{v}$ parallel and perpendicular to $\bm{Q}$ were plotted against $\nabla Q _{\perp Q}$}
\end{figure*}

\begin{figure*}[p]
\includegraphics[width=17.0cm]{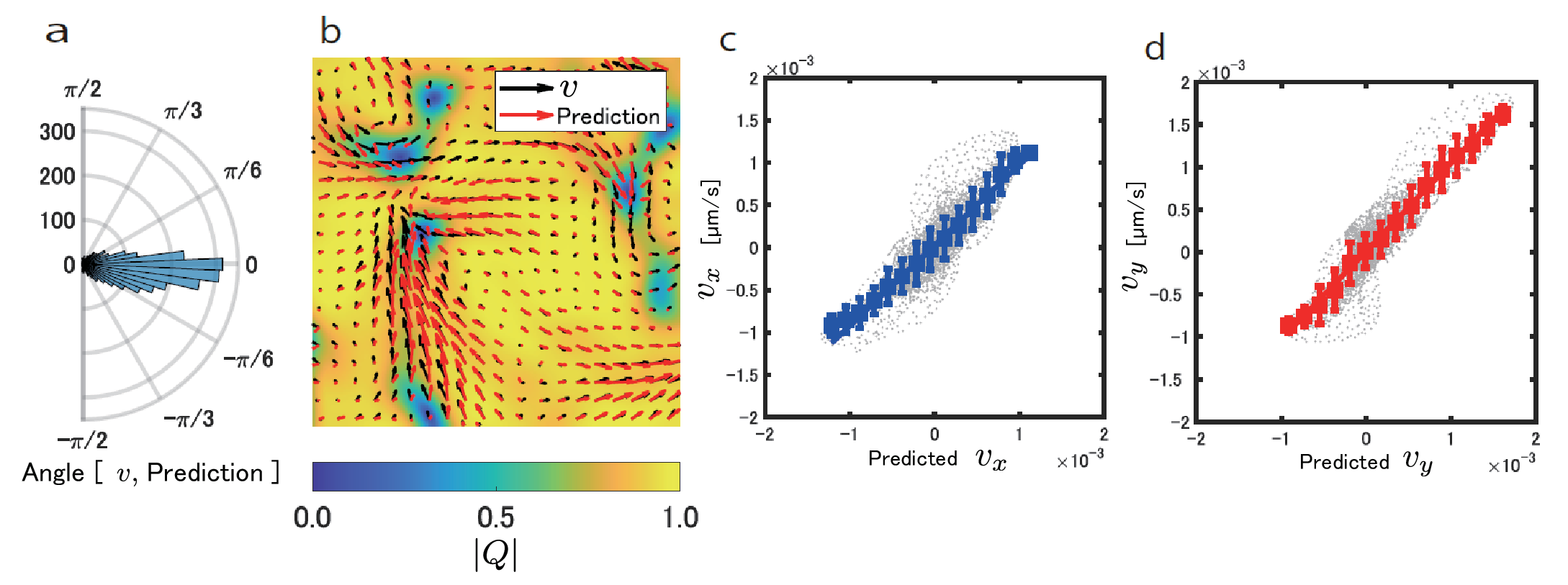}
\caption{\label{fig:Lambda_zero} Supplementary data for Fig.3 and Fig.4. 
Fitting results when $\lambda'$ is set to 0 in Eq.~(8).
a. Angle between $v$ and predicted $v$. b. Observed velocity vectors and predicted velocity vectors were plotted on the scalar order parameter $S$. c. $v_x$ compared with the prediction by Eq.~(8) without $\lambda'$. d. $v_y$ compared with the prediction by Eq.~(8) without $\lambda'$.
Symbols denote experimental results, solid lines show the linear fitting. Error bars are standard deviations for binned data sets.}
\end{figure*}

For SK-LMS-1 cells, the traction force of isolated cells was measurable. The strength $lf$ of the isolated force dipoles was then calculated as
\begin{eqnarray}
    lf &=& \sum_{\delta\bm{r}} -dA\left(\bm{T}(\bm{r}_c + \delta\bm{r})\cdot\bm{D}(\delta\bm{r})\right),\\
    \bm{D}(\delta\bm{r}) &=& |\delta\bm{r}| (\cos\theta_{{\rm dipole}}(\delta\bm{r}), \sin\theta_{{\rm dipole}}(\delta\bm{r}))^T,~\\
    \theta_{{\rm dipole}}(\delta\bm{r}) &=& -arg(\delta\bm{r}) + 2\theta_{{\rm cell}} ,
\end{eqnarray}
where $\bm{r}_c$, $\theta_{{\rm cell}}$ is the manually inputted center and major axis direction of the cell, $\delta\bm{r} = \bm{r} - \bm{r}_c$, and $dA$ is the area of the grid on which the traction force $\bm{T}$ was calculated [Fig.~\ref{fig:singleSKTFM}(a,b)]. By this calculation, we obtained $lf\sim10^4~ {\rm nN\cdot\mu m}$ [Fig.~\ref{fig:singleSKTFM}(c)]. Since this $\zeta$ is also much smaller than the value ($\zeta \sim4\times10^5 {\rm~Pa\cdot\mu m}$) measured in our traction force experiment for collective cells, it is implied that the collective effect is present also for SK-LSM-1 cells.

\bibliography{main.bib}